\documentclass{pasj00}

\begin{document}

\include{defn}
\def\cm{{\rm\thinspace cm}}
\def\gm{{\rm\thinspace gm}}
\def\dyn{{\rm\thinspace dyn}}
\def\erg{{\rm\thinspace erg}}
\def\eV{{\rm\thinspace eV}}
\def\MeV{{\rm\thinspace MeV}}
\def\g{{\rm\thinspace g}}
\def\ga{{\rm\thinspace gauss}}
\def\K{{\rm\thinspace K}}
\def\keV{{\rm\thinspace keV}}
\def\km{{\rm\thinspace km}}
\def\kpc{{\rm\thinspace kpc}}
\def\Lsun{\hbox{$\rm\thinspace L_{\odot}$}}
\def\m{{\rm\thinspace m}}
\def\Mpc{{\rm\thinspace Mpc}}
\def\Msun{\hbox{$\rm\thinspace M_{\odot}$}}
\def\Zsun{\hbox{$\rm\thinspace Z_{\odot}$}}
\def\pc{{\rm\thinspace pc}}
\def\ph{{\rm\thinspace ph}}
\def\s{{\rm\thinspace s}}
\def\yr{{\rm\thinspace yr}}
\def\sr{{\rm\thinspace sr}}
\def\Hz{{\rm\thinspace Hz}}
\def\MHz{{\rm\thinspace MHz}}
\def\GHz{{\rm\thinspace GHz}}
\def\chisq{\hbox{$\chi^2$}}
\def\delchi{\hbox{$\Delta\chi$}}
\def\cmps{\hbox{$\cm\s^{-1}\,$}}
\def\cmpssq{\hbox{$\cm\s^{-2}\,$}}
\def\cmsq{\hbox{$\cm^2\,$}}
\def\cmcu{\hbox{$\cm^3\,$}}
\def\pcmcu{\hbox{$\cm^{-3}\,$}}
\def\pcmcuK{\hbox{$\cm^{-3}\K\,$}}
\def\dynpcmsq{\hbox{$\dyn\cm^{-2}\,$}}
\def\ergcmcups{\hbox{$\erg\cm^3\ps\,$}}
\def\ergpcmps{\hbox{$\erg\cm^{-3}\s^{-1}\,$}}
\def\ergpcmsqps{\hbox{$\erg\cm^{-2}\s^{-1}\,$}}
\def\ergpcmsqpspA{\hbox{$\erg\cm^{-2}\s^{-1}$\AA$^{-1}\,$}}
\def\ergpcmsqpspsr{\hbox{$\erg\cm^{-2}\s^{-1}\sr^{-1}\,$}}
\def\ergpcmcups{\hbox{$\erg\cm^{-3}\s^{-1}\,$}}
\def\ergpcmps{\hbox{$\erg\cm^{-1}\s^{-1}$}}
\def\ergps{\hbox{$\erg\s^{-1}\,$}}
\def\ergpspmp{\hbox{$\erg\s^{-1}\Mpc^{-3}\,$}}
\def\gpcm{\hbox{$\g\cm^{-3}\,$}}
\def\gpcmps{\hbox{$\g\cm^{-3}\s^{-1}\,$}}
\def\gps{\hbox{$\g\s^{-1}\,$}}
\def\Jy{{\rm Jy}}
\def\keVpcmsqpspsr{\hbox{$\keV\cm^{-2}\s^{-1}\sr^{-1}\,$}}
\def\kmps{\hbox{$\km\s^{-1}\,$}}
\def\kmpspmp{\hbox{$\km\s^{-1}\Mpc{-1}\,$}}
\def\Lsunppc{\hbox{$\Lsun\pc^{-3}\,$}}
\def\Msunpc{\hbox{$\Msun\pc^{-3}\,$}}
\def\Msunpkpc{\hbox{$\Msun\kpc^{-1}\,$}}
\def\Msunppc{\hbox{$\Msun\pc^{-3}\,$}}
\def\Msunppcpyr{\hbox{$\Msun\pc^{-3}\yr^{-1}\,$}}
\def\Msunpyr{\hbox{$\Msun\yr^{-1}\,$}}
\def\pcm{\hbox{$\cm^{-3}\,$}}
\def\pcmsq{\hbox{$\cm^{-2}\,$}}
\def\pcmK{\hbox{$\cm^{-3}\K$}}
\def\phpcmsqps{\hbox{$\ph\cm^{-2}\s^{-1}\,$}}
\def\pHz{\hbox{$\Hz^{-1}\,$}}
\def\pmpc{\hbox{$\Mpc^{-1}\,$}}
\def\pmpccu{\hbox{$\Mpc^{-3}\,$}}
\def\ps{\hbox{$\s^{-1}\,$}}
\def\psqcm{\hbox{$\cm^{-2}\,$}}
\def\psr{\hbox{$\sr^{-1}\,$}}
\def\kmpspMpc{\hbox{$\kmps\Mpc^{-1}$}}

\SetRunningHead{The MCG--6-30-15 \emph{Suzaku} Team}{Suzaku observations of MCG--6-30-15}
\Received{2000/12/31}
\Accepted{2001/01/01}

\title{\emph{Suzaku} observations of the hard X-ray variability of
MCG--6-30-15: the effects of strong gravity around a Kerr black hole}

\author{Giovanni \textsc{Miniutti}$^1$%
  \thanks{e-mail: miniutti@ast.cam.ac.uk}, Andrew C. \textsc{Fabian}$^1$ ,
  Naohisa  {\sc Anabuki}$^2$, Jamie {\sc Crummy}$^1$, Yasushi {\sc
Fukazawa}$^3$, \\
  Luigi {\sc Gallo}$^{4,5}$, Yoshito {\sc Haba}$^6$, Kiyoshi {\sc Hayashida}$^2$,
  Steve {\sc Holt}$^7$,  Hideyo {\sc
Kunieda}$^8$, Josefin
  {\sc    Larsson}$^1$, \\ Alex {\sc  Markowitz}$^9$, Chiho {\sc
{Matsumoto}}$^{8}$,
  Masanori {\sc Ohno}$^3$, James N. {\sc Reeves}$^{9,10}$ Tadayuki {\sc
    Takahashi}$^5$, \\ Yasuo {\sc Tanaka}$^4$, Yuichi {\sc
Terashima}$^{5,11}$,  Ken'ichi
  {\sc Torii}$^2$, Yoshihiro {\sc Ueda}$^{12}$, Masayoshi {\sc
Ushio}$^5$, \\ Shin  {\sc
    Watanabe}$^5$, Makoto {\sc Yamauchi}$^{13}$, Tahir {\sc Yaqoob}$^{9,10}$ }

\affil{1 Institute of Astronomy, University of Cambridge, Madingley
  Road, CB3 0HA Cambridge, UK}

\affil{2 Department of Earth and Space Science, Osaka University, 1-1
  Machikaneyama, Toyonaka,\\ 560-0043 Osaka, Japan}

\affil{3 Department of Physics, Hiroshima University, 1-3-1
  Kagamiyama, Higashi-Hiroshima, Hiroshima 739-8526, Japan}

\affil{4 Max-Planck-Institut f\"{u}r extraterrestrische Physik,
  Postfach 1312, Garching, Germany}

\affil{5 Institute of Space and Astronautical Science, Japan Aerospace
  Exploration Agency, Yoshinodai 3-1-1, Sagamihara, \\ Kanagawa
  229-8510, Japan }

\affil{6 Department of Astrophysics, Nagoya University, Nagoya
  464-8602, Japan}

\affil{7 F. W. Olin College of Engineering, 1735 Great Plain Avenue,
  Needham, MA 02492, USA}

\affil{8 Department of Physics, Nagoya University, Furo-cho, Chikusa,
  Nagoya 464-8602, Japan}

\affil{9 Exploration of the Universe Division, Code 662, NASA Goddard
  Space Flight Center, Greenbelt, MD 20771, USA}

\affil{10 Department of Physics and Astronomy, John Hopkins
  University, 3400 N Charles Street, Baltimore, MD 21218, USA}

\affil{11 Department of Physics, Ehime University, Matsuyama 790-8577,
  Japan}

\affil{12 Department of Astronomy, Kyoto University, Kyoto 606-8502,
  Japan}

\affil{13 Department of Applied Physics, University of Miyazaki, 1-1,
  Gakuen-Kibanadi-Nishi, Miyazaki 889-2192, Japan}


%


\maketitle

\begin{abstract}
\emph{Suzaku} has, for the first time, enabled the hard X-ray
variability of the Seyfert 1 galaxy MCG--6-30-15 to be measured. The
variability in the 14--45~keV band, which is dominated by a strong
reflection hump, is quenched relative to that at a few keV. This
directly demonstrates that the whole reflection spectrum is much less
variable than the power-law continuum. The broadband spectral
variability can be decomposed into two components -- a highly variable
power-law and constant reflection -- as previously inferred from other
observations in the 2--10~keV band. The strong reflection and high
iron abundance give rise to a strong broad iron line, which requires
the inner disc radius to be at about 2 gravitational radii. Our
results are consistent with the predictions of the light bending model
which invokes the very strong gravitational effects expected very
close to a rapidly spinning black hole.
\end{abstract}

\section{Introduction}

The Seyfert 1 galaxy MCG--6-30-15 at $z=0.00775$ has been the centre
of much interest since a broad iron line was discovered in its X-ray
spectrum with the \emph{ASCA} satellite (Tanaka et al 1995). Iron line
emission is part of the reflection spectrum produced by the hard X-ray
power-law continuum in the source irradiating the accretion disc
(Guilbert \& Rees 1988; Lightman \& White 1988; Ross \& Fabian 1993)
and is broadened and skewed by Doppler, gravitational redshift
effects, and light aberration and bending (Fabian et al 1989). The low
energy extent of the line can reveal the inner radius of the accretion
disc and thus the black hole spin (for reviews see Fabian et al 2000;
Reynolds \& Nowak 2003 and Fabian \& Miniutti 2006). The iron
abundance in MCG--6-30-15 appears to be about 2--3 times the Solar value
(Fabian et al 2002), making the iron line particularly strong.

MCG--6-30-15 was observed several more times with \emph{ASCA} (Iwasawa et al
1996, 1999; Matsumoto et al 2003; Shih et al 2002), by \emph{BeppoSAX}
(Guainazzi et al 1999), \emph{RXTE} (Lee et al 1999; Vaughan \& Edelson 2001)
and by \emph{XMM--Newton} (Wilms et al 2001; Fabian et al 2002, 2003; Vaughan
\& Fabian 2004). All of these observations confirmed the presence and
general broad shape of the iron line. Evidence that the black hole in
MCG--6-30-15 is rapidly spinning has been obtained from the extreme
breadth of the line by Iwasawa et al (1996), Dabrowski et al (1997),
Wilms et al (2001), Fabian et al (2002) and most recently Reynolds et
al (2005), who determine a spin parameter $a=0.989$.

The  \emph{XMM--Newton} work emphasizes that much of the emission arise from smaller
radii, between $2-6$~$r_{\rm g}$ (where $r_{\rm g}=GM/c^2$). This may
help to explain the otherwise puzzling spectral variations shown by
the source. Shih et al (2002), Fabian et al (2002), Fabian \& Vaughan
(2003) and Taylor, Uttley \& McHardy (2003) found that the spectral
variability can be explained by a simple two component model
consisting of a highly variable power-law continuum (of almost fixed
spectral slope) and a much less variable reflection spectrum, which of
course includes the iron line. The reflection (and iron line) strength
do not follow the power-law continuum intensity, as expected in a
simple reflection picture. The small radius of much of the emission
can however explain this behaviour when it is recalled that as well as
strong gravitational red-shifting occurring in this region, there is
also strong gravitational light bending. This can disconnect
variations in the continuum from those of the reflection and has led
to the development of the light-bending model (Fabian \& Vaughan 2003;
Miniutti et al 2003, 2004), which is a generalization of earlier work
on the strong field regime (Martocchia \& Matt 1996; Reynolds \&
Begelman 1997).

If the source of the continuum emission (assumed to be an isotropic
emitter) changes location close to the black hole, then, even if the
continuum has a constant intrinsic luminosity, it appears to the
outside observer to change in brightness. This is just due to gravity
bending the light rays out of the line of sight by different amounts
depending upon the precise location of the source. Much of the
radiation is bent down onto the disc, so the observed reflection
intensity changes little. The two--component variability pattern and
its light bending interpretation have recently found application in
many other accreting black hole sources (Fabian et al 2004, 2005;
Miniutti, Fabian \& Miller 2004; Ponti et al 2006).

One striking feature of the model reflection spectrum is a large
reflection hump peaking at 20--40~keV. This is where the disc albedo
is highest; at lower energies the albedo is reduced by photoelectric
absorption and at higher energies it is reduced (at a given energy) by
Compton down-scattering of the photons (see e.g. George \& Fabian
1991). The presence of the Compton hump has been confirmed in the
spectrum of MCG--6-30-15 by \emph{BeppoSAX} and RXTE observations
(Guainazzi et al 1999; Lee et al 2000; Fabian et al 2002). What those
observations have not done is to determine the variability of the
Compton hump and show that it varies in the same way as the rest of
the reflection spectrum\footnote{The possible anticorrelation between
the reflection continuum and iron line reported by Lee et al (2000)
may in part be due to confusion caused by variation in the fitted
power-law index (see discussion of a spurious reflection--photon index
correlation found with RXTE data by Vaughan \& Edelson 2001).}. Here
we carry out this important step with the Hard X-ray Detector (HXD) on
\emph{Suzaku}.

\begin{figure*}
 \begin{center}
 \includegraphics[width=0.38\textwidth,height=0.9\textwidth,angle=-90]{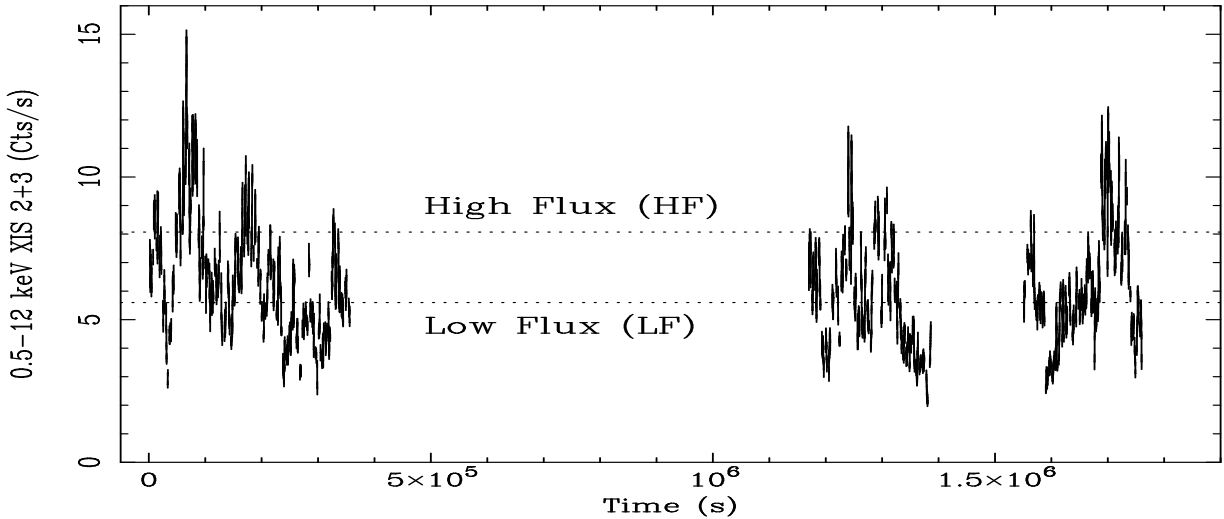}
\end{center}
\caption{The light curve from the XIS2/XIS3 detectors in the
  0.5--12~keV band. We also show the count rates chosen to select the
  High Flux (HF) and Low Flux (LF) states. They are chosen so that the
  HF and LF spectra have approximately the same number of counts
  ($6.5\times 10^5$ in the whole 0.5--12~keV band). }
\label{totallc} 
\end{figure*}

We concentrate on the \emph{Suzaku} data above 3~keV since there is a
complex warm absorber in MCG--6-30-15 (Otani et al 1996; Lee et al
2001; Turner et al 2003,2004). The difference spectrum between low and
high states of the source shows that absorption is minimal above 3~keV
(Fabian et al 2002; Turner et al 2004) and High Energy Transmission
Grating (HETG) spectra from \emph{Chandra} show no absorption feature
around 6.5~keV which would indicate absorption by species of
intermediate ionization which could affect our fits (Young et al
2005).

\section{The \emph{Suzaku} observations}

MCG--6-30-15 was observed four times by \emph{Suzaku} (Mitsuda et al
2006), once between August 17--19 in 2005 for about 45~ks and three
times in January 2006 with longer exposures. Here we focus on the
three 2006 observations performed between 9--14 (150~ks), 23--26
(99~ks)and 27--30 (97~ks) January 2006. We use event files from
revision 0.7 of the \emph{Suzaku} pipeline. Version 0.7 processing is
an internal processing applied to the \emph{Suzaku} data obtained
during the Suzaku Working Group phase, for the purpose of establishing
the detector calibration as quickly as possible. Some processes that
are not critical for most of the initial calibration and scientific
studies, e.g., aspect correction, fine tuning of the event time
tagging of the XIS data, are skipped in version 0 processing, making
the quality of the products limited in these directions compared with
the official data supplied to guest observers.  The XIS data were
screened with {\tt XSELECT} as standard (see e.g. Koyama et al 2006).
The XIS products were extracted from circular regions of 4.3' radius
centred on the source, while background products were extracted from
two smaller circular regions offset from the source (and avoiding the
chip corners with calibration sources) with a total area equal to that
of the source region. The latest response and ancillary response files
provided by the instrument teams were used. For the HXD/PIN (Takahashi
et al 2006), instrumental background spectra were extracted from time
dependent models provided by the HXD instrument team, based upon a
database of non X-ray background observations made by the PIN diode to
date. Since the background modeling is the key issue for the hard
X-ray measurement with the HXD, the HXD team has provided two
independent background models, model A and B, which use different
algorithms (Kokubun et al. 2006). Spectral analysis of the source
spectrum using the two models was found to give statistically
indistinguishable results for the three 2006 January MCG--6-30-15
observations over the whole 14--45~keV band used here, and we use the
background model A in our analysis. The response files appropriate for
the XIS nominal position observation were chosen dated as of
2006/08/07.

We have extracted products for the three front--illuminated CCD XIS
detectors (XIS0, XIS2, and XIS3) and for the back--illuminated CCD (XIS1).
The XIS2 and XIS3 detectors are found to produce remarkably similar
spectra in the whole band used here. The XIS0 spectrum is slightly
flatter than that from the XIS2 and XIS3 detectors so we proceed by
co--adding just the XIS2 and XIS3 products in our analysis (see Yaqoob
et al 2006 for mention of structures in the XIS0 and XIS1 spectra).  In
Fig.~\ref{totallc}, we show the broad-band 0.5--12~keV background
subtracted light curve from the XIS2 and XIS3 detectors during the three
pointed observations in 2006. We also show, as a reference, the count
rate levels selected to define the High Flux (HF) and Low Flux (LF)
states which will be used to study the spectral variability of
MCG--6-30-15. As normal for this source (thought to harbour a black
hole with a mass of $\sim 3\times 10^6~M_\odot$, McHardy et al 2005)
the light curve exhibits large amplitude and relatively short
timescale variability with variations up to factors 2--3 in a few ks.

\section{The 3--12~keV XIS spectrum}

\begin{figure}
 \begin{center}
{
 \includegraphics[width=0.33\textwidth,height=0.42\textwidth,angle=-90]{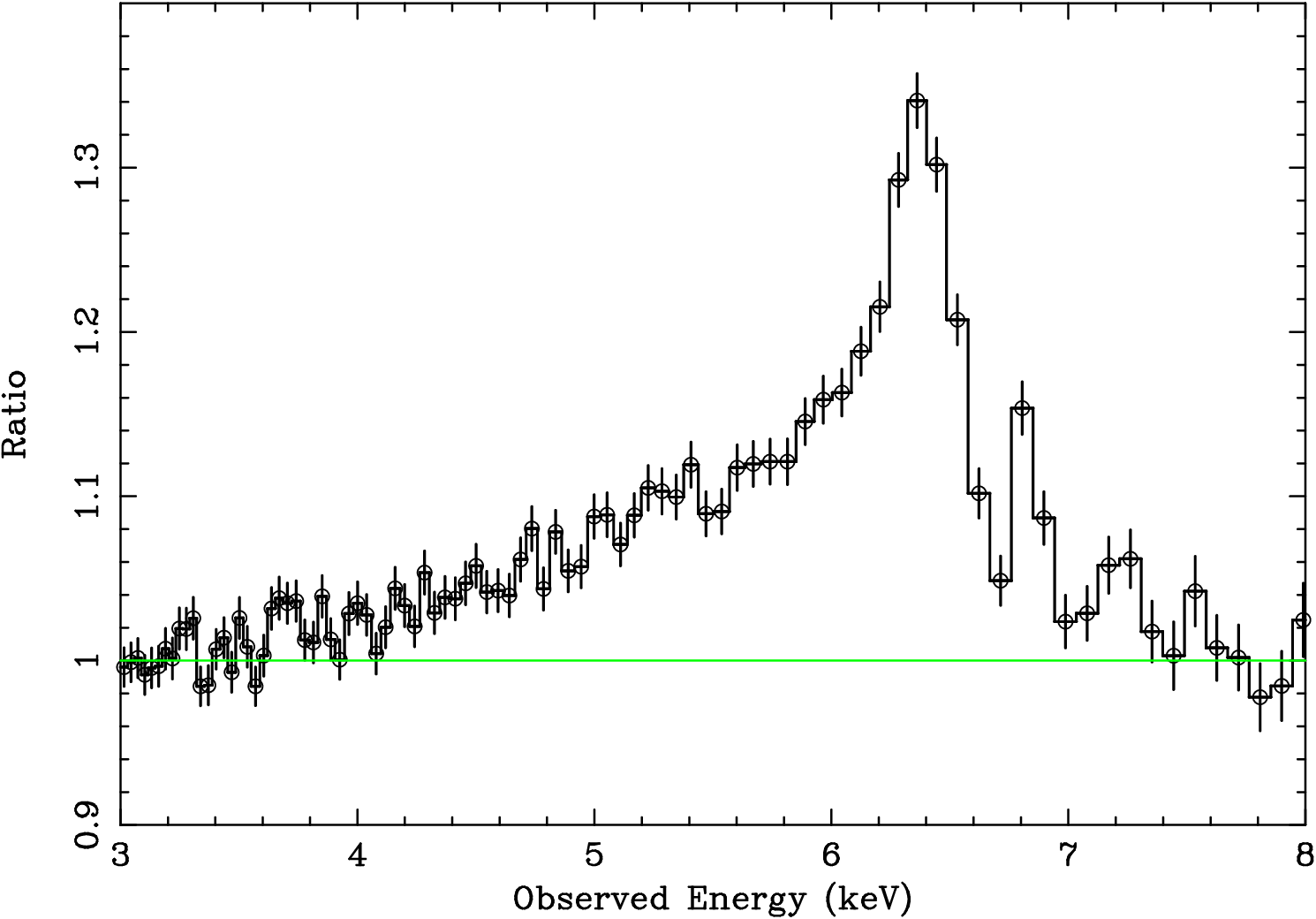}
{\vspace{0.2cm}}
 \includegraphics[width=0.33\textwidth,height=0.42\textwidth,angle=-90]{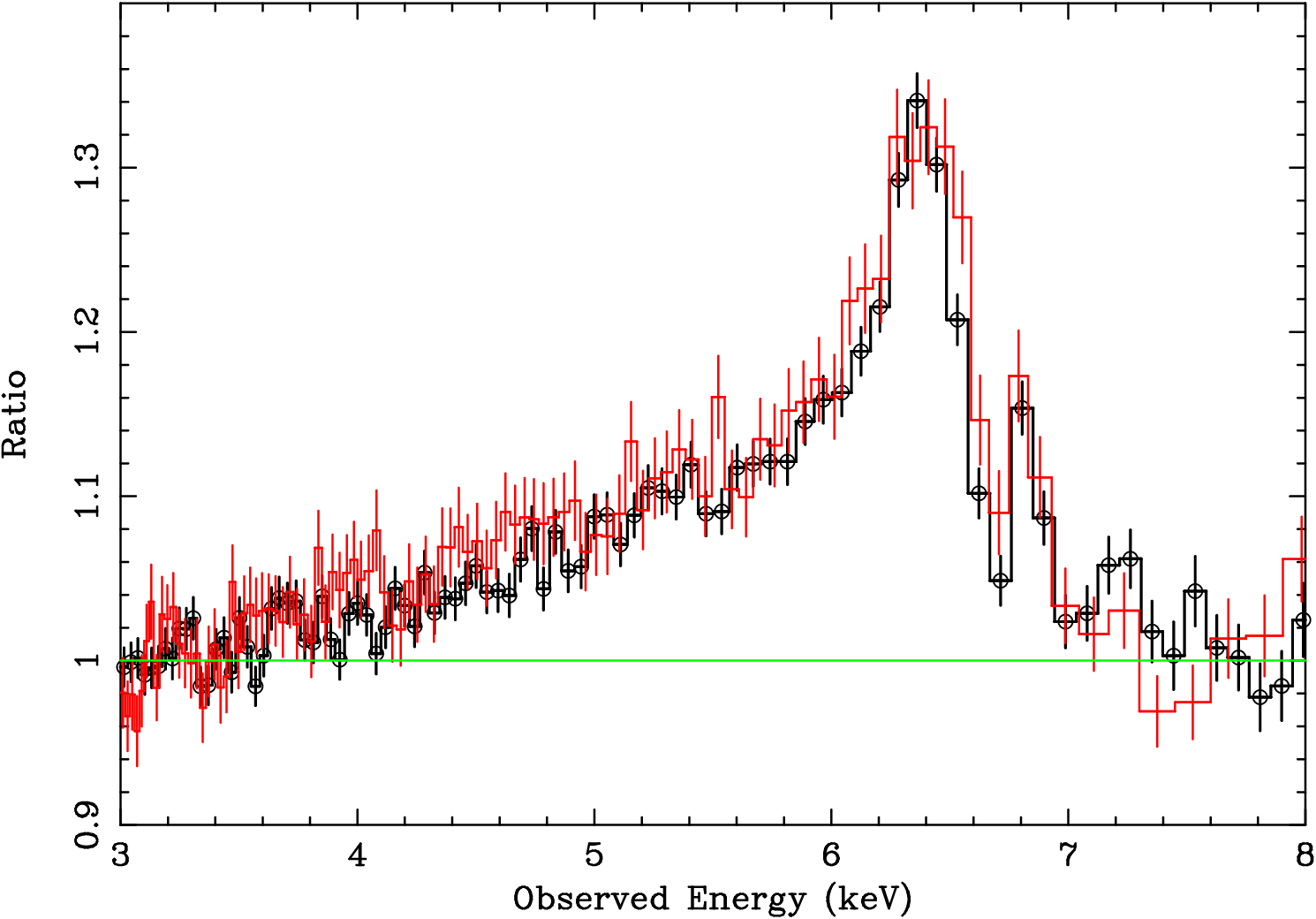}
}
\end{center}
\caption{{\bf Top:} The \emph{Suzaku} XIS data divided by a power law
  model fitted in the 3--4~keV and 7.5--10~keV band are shown in the
  most relevant 3--8~keV band. {\bf Bottom:} The data from the \emph{XMM--Newton}
  EPIC--pn detector (red) are superimposed on the XIS
  data (black). In both cases the model is a power law absorbed by the
  Galactic column density and fitted in the 3--4~keV and 7.5--12~keV
  band.}
\label{Feprofiles}
 \end{figure}

We start our analysis of the \emph{Suzaku} data by considering the
3--12~keV time--averaged co--added spectrum from the XIS2 and XIS3
front--illuminated CCD detectors. (XIS response and ancillary files
20060213.rmf and 20060415.arf, with a 6~mm extract radius, were used.)
The most important feature in this energy band is the strong, skewed,
relativistic Fe K$\alpha$ line which characterizes the X--ray spectrum
of MCG--6-30-15 enabling us to explore the nature and geometry of the
accretion flow close to the central black hole with much higher
accuracy than in any other object so far. In Fig.~\ref{Feprofiles} we
show the ratio of the data to a simple power law model fitted in the
3--4~keV and 7.5--12~keV band and absorbed by the Galactic column
density ($4.08\times 10^{20}$~cm$^{-2}$). The residuals clearly show
the asymmetric and broad Fe K$\alpha$ line profile (top panel). In the
bottom panel, we superimpose the Fe K$\alpha$ profile as observed with
the \emph{XMM--Newton} EPIC--pn camera in 2001. The agreement between
the two instruments is remarkable and demonstrates the excellent level
of the \emph{Suzaku} XIS calibration even in the early stages of the
mission. The 2--10~keV flux is $4.1\times
10^{-11}$~erg~cm$^{-2}$~s$^{-1}$ in the \emph{XMM--Newton} observation
and $4.0\times 10^{-11}$~erg~cm$^{-2}$~s$^{-1}$ in the \emph{Suzaku}
one. Notice the absorption/emission structures in the blue wing of the
relativistic line which are clearly detected by both the EPIC--pn and
the XIS detectors with excellent agreement.

\subsection{The Fe K band and the relativistic line}

As a first attempt to fit the Fe K$\alpha$ line profile we consider
the simplest possible spectral model comprising a power law continuum
absorbed by a column of neutral matter (fixed at the Galactic value
$N_H=4.08\times 10^{20}$~cm$^{-2}$) and a set of Gaussian emission
lines. We consider first three Gaussian emission lines: a narrow
unresolved $\sim$6.4~keV emission line (the narrow component of the Fe
line from distant matter), two narrow unresolved $\sim$6.7~keV and
$\sim$6.97~keV absorption lines (the Fe\textsc{xxv} and
Fe\textsc{xxvi} resonant absorption line already detected by
\emph{Chandra}, Young et al. 2005), and an additional Gaussian
emission line with width free to vary (representing the broad
relativistic Fe line). We obtain a fit with $\chi^2=2550$ for 2235
degrees of freedom (dof). The narrow component of the Fe line is at
$6.43^{+0.01}_{-0.02}\keV$ (Uncertainties are quoted throughout the
paper at the 90 per cent confidence level) and has an equivalent width
(EW) of only $30\pm 5$~eV. Absorption lines are detected at $6.74\pm
0.03$~keV and $7.04\pm 0.05$~keV. The broad Fe line is at $5.88\pm
0.05\keV$, has a width of $840\pm 40$~eV and an EW of $305\pm 20$~eV.

However, the best--fitting model leaves clear residuals in the
3--12~keV band. In particular the broad Fe line is not properly
modelled and the residuals suggest the introduction of a double
Gaussian model with one broad Gaussian around 6.4~keV and an even
broader one at lower energies to model the extended red wing. We
obtain a significant improvement of the statistics with $\chi^2=2398$
for 2232 (Table~1).  The energies of the three unresolved lines are
$6.41^{+0.03}_{-0.02}$~keV for the narrow Fe emission line, and
$7.04\pm 0.05$~keV, and $6.73\pm 0.04$~keV for the absorption lines.
Their equivalent widths are $25\pm 5$~eV, $-(12\pm 8)$, and $-(15\pm
6)$~eV respectively. The upper limit on the EW of the narrow Fe
K$\alpha$ emission line (only 30~eV) indicates that reflection from
distant matter plays a minor and marginal role in MCG--6-30-16, as
already demonstrated by high--resolution spectroscopy with the HETG
\emph{Chandra} gratings (Lee et al 2002; Young et al 2005). The two
absorption line energies are slightly higher than the rest--frame
energies of Fe\textsc{xxv} and Fe\textsc{xxvi} resonant absorption and
are consistent with an origin in a common outflow with velocity of few
thousand km~s$^{-1}$. In particular, they are both consistent in
equivalent width and outflow velocity with the values measured with
\emph{Chandra} ($2.0^{+0.7}_{-0.9}\times 10^3$~km~s$^{-1}$, Young et
al 2005). The photoelectric absorption implied by such a thin highly
ionized absorber is small and has a negligible effect on the fits
described here.

As for the broad Fe line, the parameters of the two broad Gaussian
lines describing its profile are reported in Table~1 (first model
``Double Gaussian''). They are both clearly resolved and their
cumulative EW is $320\pm 45$~eV, although this value should be taken
with care since the continuum is a simple absorbed power law and does
not include the reflection continuum which must be associated with
the broad Fe line. It is interesting to note that the width of the Fe
line core at $\sim 6.45$~keV is $\sigma > 260$~eV (i.e.  FWHM$\simeq
31950$~km~s$^{-1}$) suggesting that it originates in the outer disc at
$< 80~r_g$ from the central black hole, while the red wing has a width
$>760$~eV indicating that the emitting matter is located within
$6.5~r_g$ from the centre, already implying, even by means of a very
simple and phenomenological model, that the black hole in MCG--6-30-15
is most likely a spinning Kerr black hole in which the accretion disc
extends down within the marginal stable orbit for a non--rotating
Schwarzschild black hole ($6~r_g$).

\subsection{A self--consistent reflection model}

The multiple Gaussian fit described above is merely a phenomenological
description of the hard spectrum. The Gaussian emission lines and,
most importantly, the broad Fe line are the clear signature of X-ray
reflection and the above spectral model did not include any reflection
continuum. Here we build a much more self--consistent model in which
the broad Fe line is computed together with the associated reflection
continuum. We use a grid of models from Ross \& Fabian (2005) to
obtain the X--ray reflection spectrum. The X--ray reflection model has
the spectral slope of the illuminating power law $\Gamma$, the
ionization parameter $\xi$, the Fe abundance, and the normalization as
free parameters. However, we forced the photon index to be the same as
the power law continuum for consistency. Since the reflection model
does not include Ni as an element, we also include a Ni K$\alpha$ line
with energy fixed at 7.47~keV.

In order to reproduce the relativistic broad Fe line profile, the
reflection spectrum is convolved with a relativistic kernel derived
from the Laor (1991) code. The relativistic blurring parameters are
the emissivity indexes $q_{\rm in}$ and $q_{\rm out}$ (where the
emissivity is $\epsilon = r^{-q_{\rm in}}$ within the innermost
$6~r_g$ and $\epsilon = r^{-q_{\rm out}}$ outside), the inner disc
radius $r_{\rm in}$, and the observer inclination $i$. The outer disc
radius is fixed at its maximum allowed value of $400~r_g$.  The choice
of a broken power law emissivity profile is motivated by the previous
long \emph{XMM--Newton} observation of MCG--6-30-15 (Vaughan \& Fabian
2004).

\begin{table*}
\begin{center}
  \caption{Results of spectral fits to the 3--12~keV XIS2 and XIS3
    time--averaged co--added spectrum with the different models used
    to describe the relativistic Fe line of MCG--6-30-15. We present
    results for a phenomenological double--Gaussian fit and for a much
    more self--consistent relativistically blurred reflection model.
    For the reflection model, we measure an Fe abundance of
    $2.0^{+1.4}_{-0.6}$ times solar. We only report results for the
    relativistic Fe line here. A more detailed fit is presented in
    Table~2, where high--energy data from the HXD/PIN detector are
    also included. A
    subscript $_p$ indicates that the parameter 
    reached its min/max allowed value.
    }
\begin{tabular}{cccccccc}
  \hline
  \multicolumn{8}{l}{{\bf{Double Gaussian}}}   \\
  \hline
  \\
  \multicolumn{1}{c}{Continuum} & \multicolumn{3}{c}{K$\alpha$
    Red Wing} &\multicolumn{3}{c}{K$\alpha$ Blue Core}
  & $\chi^{2}/dof$ \\
  \\
  $\Gamma$ & $E$ & $\sigma$ & $EW$ & $E$ & $\sigma$ & $EW$
  & \\
  $1.96\pm 0.02$ & $5.38\pm 0.10$ & $840^{+70}_{-80}$ &
  $130 \pm 15$ & $6.45^{+0.02}_{-0.05}$ & $290\pm 30$ &
  $190\pm 30$ &  2398/2232
  \\ \\
  \hline
  \multicolumn{8}{l}{{\bf{Blurred Reflection}}}   \\
  \hline
  \\
  \multicolumn{1}{c}{Continuum} & \multicolumn{4}{c} {Relativistic
    Blurring} & \multicolumn{2}{c}{Reflector}  &$\chi^{2}/dof$ \\
  \\
  $\Gamma$ & $i$ &  $r_{\rm{in}}$ & $q_{\rm{in}}$ & $q_{\rm{out}}$ & $R$ &  $\xi
$ &
  \\
  $2.18^{+0.07}_{-0.06}$ & $38\pm 4$ & $1.6^{+0.6}_{-0.365p}$ & $4.6^{+0.6}_{-0.
9}$ &
  $2.6\pm 0.3$  & $2.8\pm 0.9$ & $65\pm 45$ & 2360/2230
  \\
  \\
  \hline
\label{tab1}
\end{tabular}
\end{center}
\end{table*}

The model is applied to the co--added XIS2 and XIS3 time--averaged
spectrum in the 3--12~keV band and we keep the three unresolved
Gaussian lines as above. The results of the spectral fitting are
reported in Table~\ref{tab1} (``Blurred Reflection'' model). We obtain
a better fit with $\chi^2=2360$ for 2230 dof. The better fit with
respect to the phenomenological Double Gaussian model described above
is due to a better description of the overall relativistic Fe line
profile. In addition, the model accounts for some high--energy
residuals above about 10~keV which were present in the above
best--fitting solution due to the unmodelled reflection component.

We measure a relatively standard $\Gamma=2.18^{+0.07}_{-0.06}$
continuum slope which compares very well with previous results with
\emph{XMM--Newton}. The reflection component contributes significantly
to the hard spectrum of MCG--6-30-15 and we measure a reflection
fraction of $R=2.8\pm 0.9$ ($R=1$ corresponds to the level of
reflection expected from $2\pi$~sr), fully consistent with previous
results from a simultaneous \emph{XMM--Newton} and \emph{BeppoSAX}
observation (Vaughan \& Fabian 2004). The precise value of the
reflection fraction will be better constrained in a subsequent
analysis when the high energy data from the HXD/PIN detector are
considered as well. The reflector is only mildly ionized with $\xi =
65\pm 45$~erg~cm~s$^{-1}$ and the inclination is constrained to be
$38^\circ \pm 4^\circ$.

As for the relevant relativistic blurring parameters, the XIS data are
able to constrain the inner disc radius to be smaller than $r_{\rm in}
< 2.2~r_g$. If, as it is customary, $r_{\rm in}$ is identified with
the innermost stable circular orbit around a Kerr black hole, such a
small 90 per cent upper limit implies that the black hole spin in
MCG-6-30-15 is $a > 0.917$, i.e. the black hole is an almost maximally
spinning Kerr black hole. The inner and outer emissivity profiles are
not consistent with each other within the errors and $q_{\rm in}
\simeq 4.8$ is steeper than $q_{\rm out} \simeq 2.6$. However, if the
two indices are forced to be one and the same, we obtain $q_{\rm in}
\equiv q_{\rm out} = 3.1^{+0.5}_{-0.3}$ with only a slight worsening
of the fit statistic ($\Delta\chi^2=6$).

\section{Adding the HXD/PIN data; the 3--45~keV spectrum}

\begin{figure}
 \begin{center}

\includegraphics[width=0.33\textwidth,height=0.42\textwidth,angle=-90]{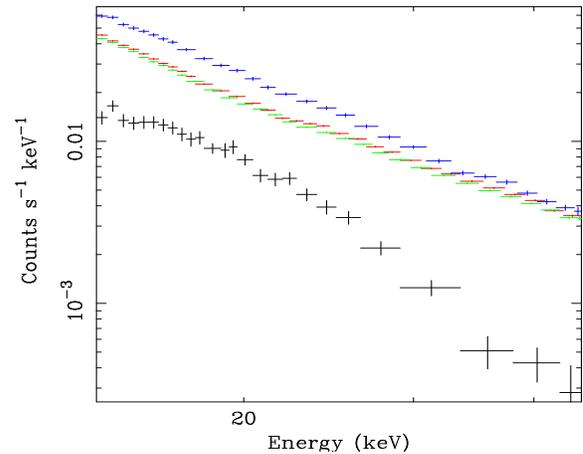}
\end{center}
\caption{PIN data displayed as total countrate (blue, upper),
backgrounds
A and B (red and green, respectively, middle) and source 
(black lower) spectra.}
\label{PINspectra}
 \end{figure}

In the above analysis the relativistic Fe line is associated with its
own reflection continuum and a self--consistent model to the 3--12~keV
spectrum is found. However, one of the main characteristics of the
\emph{Suzaku} mission is the presence of the HXD PIN hard X-ray
detector providing high quality data above 12~keV
(Fig.~\ref{PINspectra}). The spectral analysis of the XIS data
revealed the presence of a strong reflection component associated with
the broad relativistic Fe line which would produce a strong Compton
hump around 20--30~keV. The HXD data above 12~keV are thus crucial to
investigate the reflection component further and will enable us to
infer the reflection parameters and to measure the direct and
reflection continua with high accuracy. 

\begin{figure}
 \begin{center}
 \includegraphics[width=0.33\textwidth,height=0.42\textwidth,angle=-90]{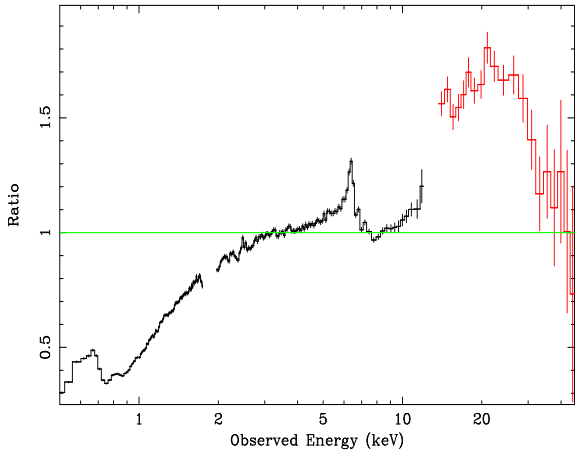}
\end{center}
\caption{The XIS2/XIS3 and PIN data in the 0.5--45~keV band are
  compared with a simple power law fitted in the 3--45~keV band
  ignoring the Fe K and Compton hump energy bands (4-7.5~keV and
  14-30~keV). The ratio plot shows the main spectral components which
  are the power law continuum, absorption below about 3~keV, and a
  strong smeared reflection component comprising the relativistic
  broad Fe line and Compton hump. Data have been re-binned for
  visual clarity.}
\label{XISandPINPLfit}
 \end{figure}

 In Fig.~\ref{XISandPINPLfit}, we show the XIS and HXD/PIN
 time--averaged spectrum in the 0.5--45~keV band. The Figure shows the
 data to model ratio for a power law model fitted above 3~keV and
 ignoring the bands where reflection is expected to dominate. The
 ratio clearly shows the presence of both the relativistic Fe line in
 the XIS data and also the presence of a large Compton hump around
 20~keV visually confirming that the hard spectrum of MCG--6-30-15 is
 dominated by a strong reflection component. Absorption is present
 affecting the soft band below 3~keV. We avoid the complication of
 modelling the complex warm absorber and we concentrate, in this first
 paper on the \emph{Suzaku} observation of MCG--6-30-15, on the data
 above 3~keV. A detailed model of the warm absorber will be presented
 elsewhere where the data from the back--illuminated (more sensitive
 at soft energies) XIS detector will be also considered.

We consider our best--fitting spectral model to the 3--12~keV XIS data
and extend it to the PIN data between 14~keV and 45~keV. We include a
cross-normalization constant between the XIS and PIN data and find
that it is constrained to be $1.10\pm 0.05$ (where the XIS constant is
set to $1$). This factor is consistent with the results obtained from
Crab observations (Kokubun et al. 2006). Since the available HXD/PIN
background model A does not include the contribution from the X--ray
cosmic background, a spectral model of the form $2.06\times 10^{-6}
(E/100~keV)^{-1.29} {\mathrm{exp}}(-E/41.13~keV)$ was included in all
spectral fits in the HXD/PIN band (the model being valid up to about
70~keV). The model is based on the \emph{HEAO}--A1 spectrum,
re--scaled to account for the HXD field of view.
 
 The best--fitting XIS model provides an acceptable description of the
 3--45~keV data and we obtain a good quality fit of $\chi^2=2495$ for
 2312 dof. The model parameters are very much consistent with those
 obtained from the 3--12~keV parameters. In fact, extending the model
 up to 45~keV provides a reasonable description of the hard data,
 reproducing most of the large Compton hump seen in
 Fig.~\ref{XISandPINPLfit}, even before fitting. The results for the
 Time--Averaged (TA) spectrum in the 3--45~keV band are reported in
 Table~2 together with similar fits to the High Flux (HF) and
 Low Flux (LF) spectra which will be discussed below. The reflection
 component contributes significantly to the total flux and we measure
 a large reflection fraction of $3.8\pm 0.7$. As a reference, and
 since it will be useful in a spectral variability analysis below, we
 report here that its contribution represents the $33 \pm 6$ per cent
 of the total flux in the 3--12~keV band and the $51\pm 10$ per cent
 in the 14--45~keV band. The relativistic blurring parameters are
 consistent within the errors with those already reported in
 Table~\ref{tab1} and the same is true for the continuum power law
 slope and for the ionization parameter.

\begin{figure}
 \begin{center}
 \includegraphics[width=0.33\textwidth,height=0.42\textwidth,angle=-90]{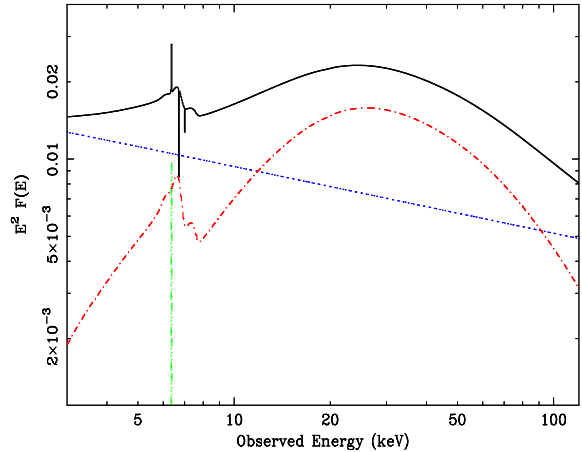}
\end{center}
\caption{The best--fit model components to the time--averaged spectrum
of MCG--6-30-15.}
\label{Suzakumo}
 \end{figure}

 As a consistency check, we have also determined the level of the
 reflection component by using the spectrum above 7.5~keV, so avoiding
 the iron line. For this the {\tt pexrav} model in XSPEC was used,
 with results $\Gamma=2.2$, reflection fraction $R=2-4$ and iron
 abundance poorly constrained ($A_{\rm{Fe}}=0.5-2$ times solar). We
 note that such reflection predicts a {\em strong} iron line with an
 equivalent width of at least 300~eV (taking conservatively an iron
 abundance of unity and reflection fraction of 2), using the
 predictions of George \& Fabian (1991). The limit on any narrow
 component to the iron emission feature is at least ten times smaller
 than this (this work; Lee et al 2002; Young et al 2005), thereby
 ruling out any non-smeared origin for the reflection component. The
 reflection must be due to matter well within the optical broad line
 region. We also stress that if the evidence for relativistic
 reflection is taken into account and the {\tt pexrav} component is
 blurred accordingly, the shape of the reflection continuum is broader
 and redshifted requiring a higher Fe abundance ($A_{\rm{Fe}}=1.5-3.5$
 times solar) and reflection fraction ($R=3-5$) to fit the data,
 while the continuum photon index is basically unaffected. The blurred
 and Fe overabundant reflector described by {\tt pexrav} is thus fully
 consistent with our best--fitting results with a much more complex
 spectral model including the relativistic Fe line (see Table~2).

\begin{figure}
 \begin{center}
{
 \includegraphics[width=0.33\textwidth,height=0.42\textwidth,angle=-90]{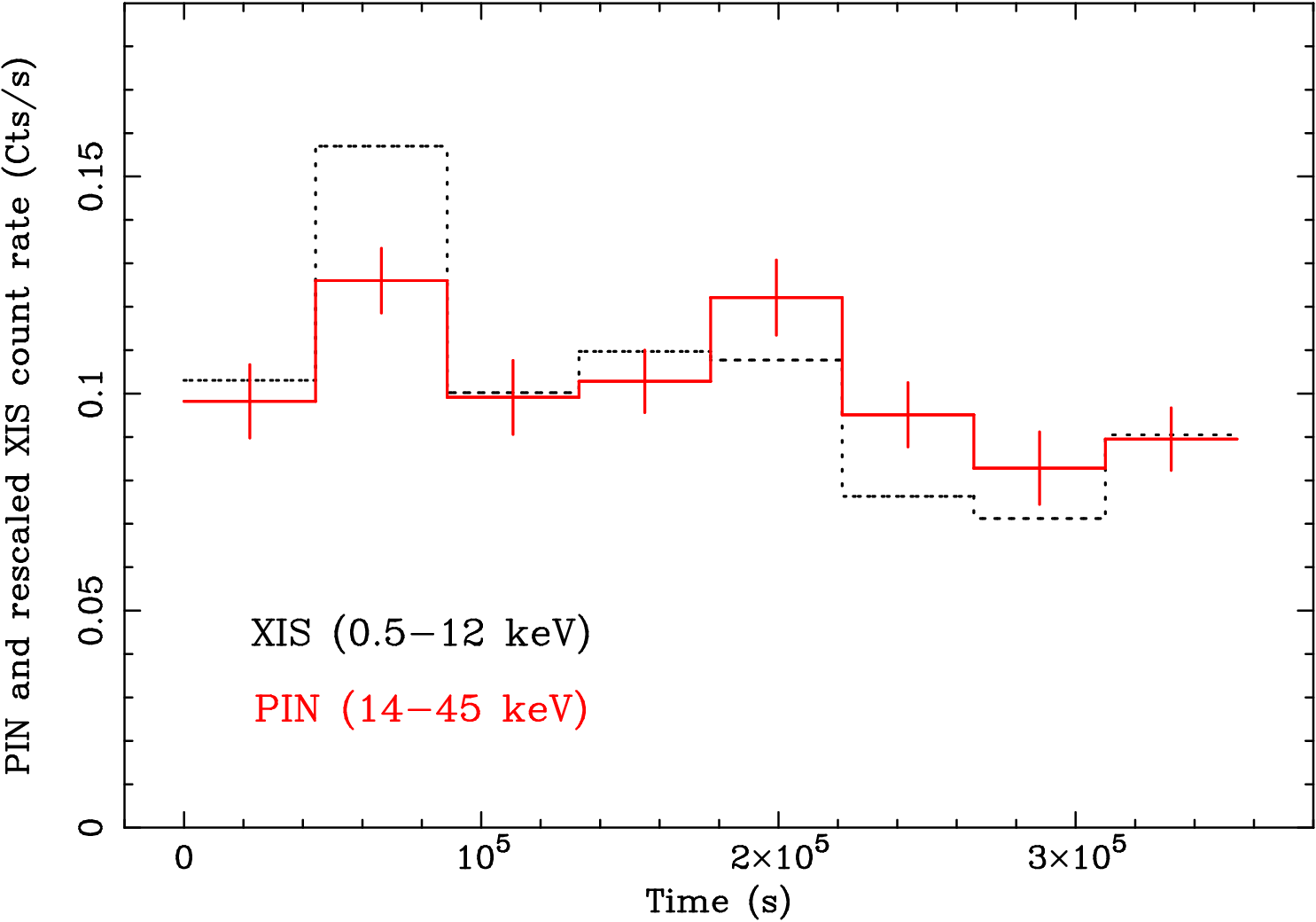}
{\vspace{0.2cm}}
 \includegraphics[width=0.33\textwidth,height=0.42\textwidth,angle=-90]{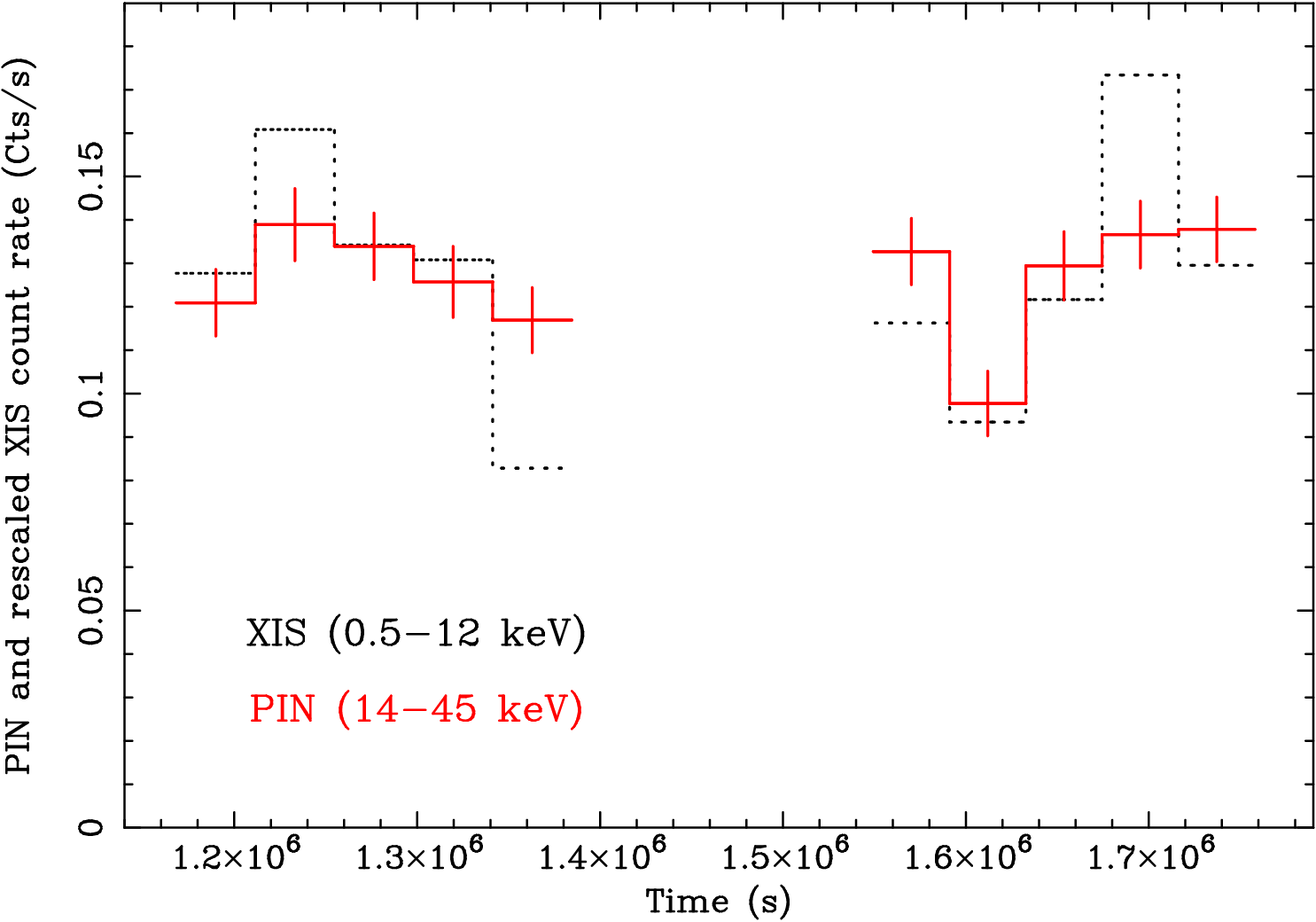}
}
\end{center}
\caption{The background subtracted light curve of MCG--6-30-15 during
  the three 2006 pointed observations (first in the top panel, second
  and third in the bottom panel). The dotted black light curve is the
  XIS2 and XIS3 light curve rescaled to the mean HXD/PIN count rate in
  each orbit. In red, we show the HXD/PIN light curve. The time bin is
  about 45~ks. Time elapses from the start of the first observation.}
\label{45kslc}
 \end{figure}

\section{Flux and spectral variability}

 The HXD/PIN detector has no imaging capabilities and thus there is no
simple way to obtain a background subtracted light curve by selecting
appropriate regions. However, background subtracted light curves can
be obtained by extracting spectra in each time interval, subtracting
the background from the available models in the same time interval,
and reading out the background subtracted count rate.  In background
model A, the background of the HXD is calculated from the data base by
using the count rate of the upper discriminator, which has a strong
correlation with the flux of the non-X-ray background. The procedure
must also include a model for the cosmic X--ray background (not
included in the available background models) and we use the same model
as for spectral fitting by simulating the cosmic X--ray background
HXD/PIN spectrum and adding it to the instrumental background. We have
applied the above procedure to the HXD/PIN data by selecting, after
some experimentation, a timescale of 45~ks enabling us to obtain about
5000 background subtracted counts per time interval in the 14--45~keV
PIN band on average. In Fig.~\ref{45kslc} we show the XIS2 and XIS3 and
the HXD/PIN light curves for the three 2006 observations in the
0.5--12~keV and 14--45~keV respectively. To ease the comparison, the
XIS light curves have been re--scaled so that they have the same mean
count rate as the PIN ones in each segment (i.e. observation). It is
visually clear that the harder band (solid red) has a much lower
variability amplitude than the softer one (dotted black). In the
following we explore a set of different techniques with the aim of
exploring the flux and spectral variability of the source.

\subsection{Flux--flux plots}
\begin{figure}
 \begin{center}
{
 \includegraphics[width=0.33\textwidth,height=0.42\textwidth,angle=-90]{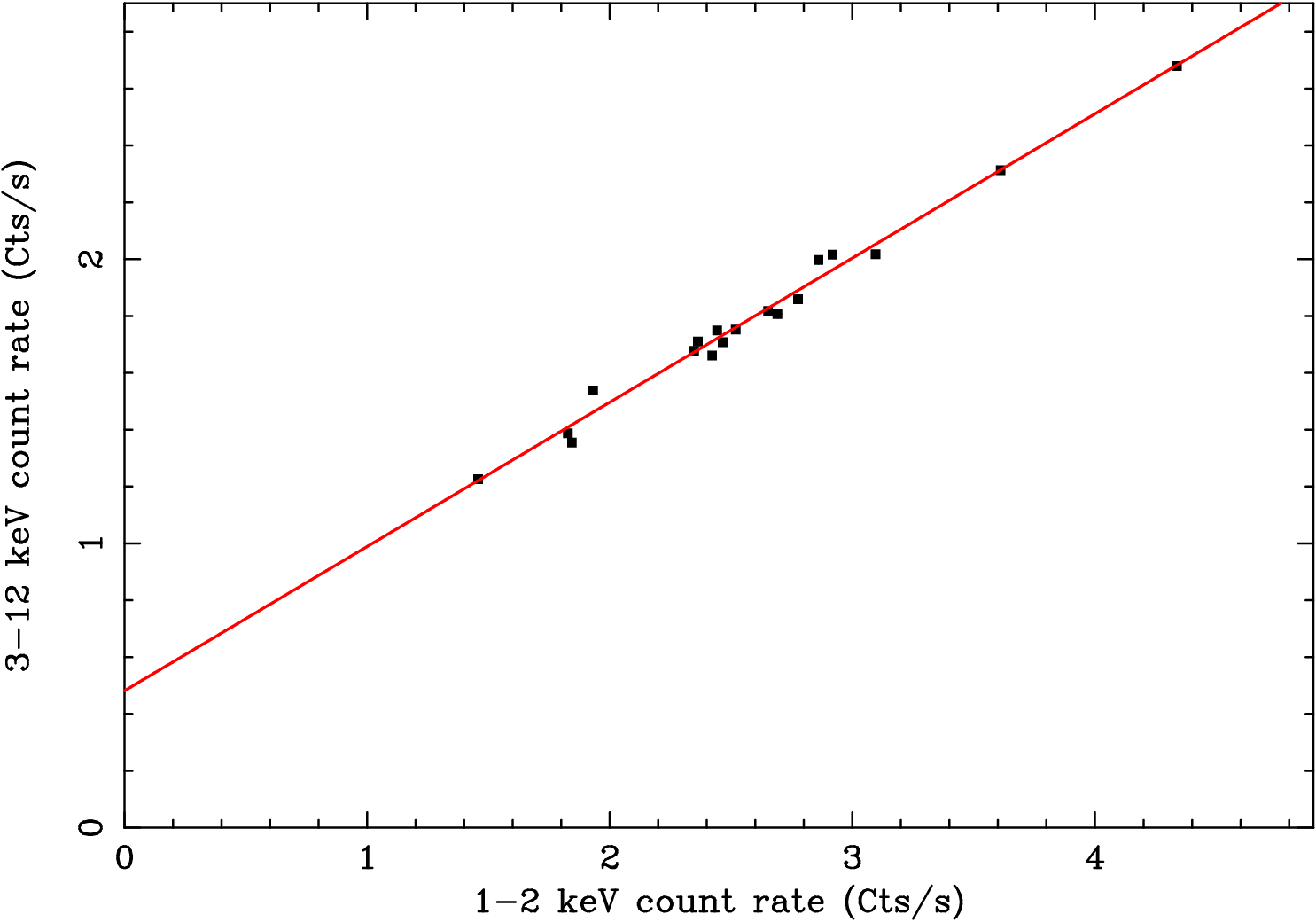}
{\vspace{0.2cm}}
 \includegraphics[width=0.33\textwidth,height=0.42\textwidth,angle=-90]{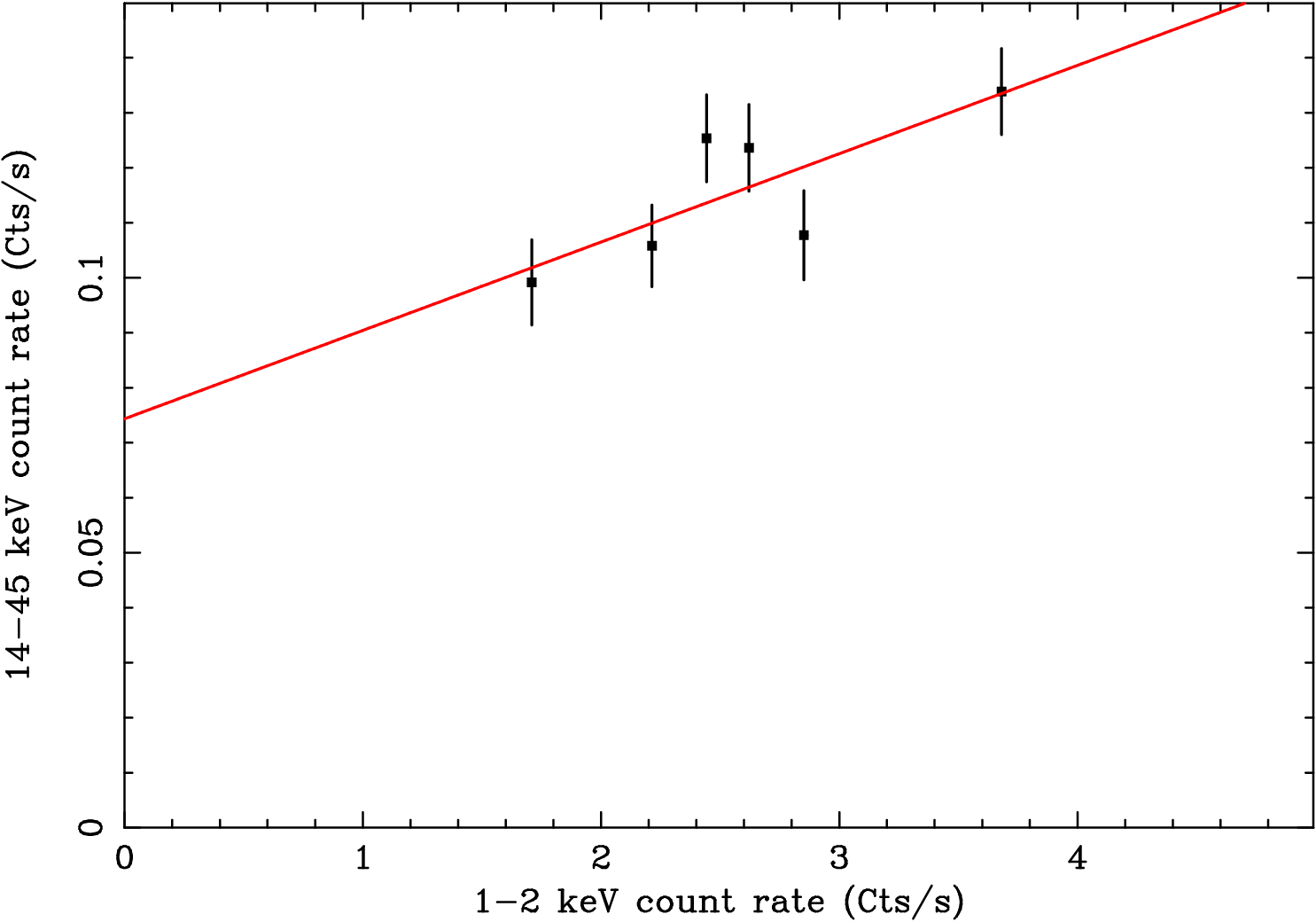}
}
\end{center}
\caption{{\bf{Top:}} The XIS2 and XIS3 count rate in the hard 3--12~keV
  band is plotted against that in the 1--2~keV band from the 45~ks
  light curve. The relationship is linear and reveals a clear hard
  off--set.  {\bf{Bottom:}} Flux--flux plot for the 14--45~keV HXD/PIN
  data against the 1--2~keV reference band. Due to larger spread in
  the PIN band (see text for details), we have re-binned the original
  flux--flux plot according to the 1--2~keV flux so that each
  data point shown here comprises three original data points. The
  binned flux--flux plot exhibits a linear relationship, although some
  scatter is present.}
\label{ff}
 \end{figure}

As a first model--independent way to characterize the flux/spectral
variability of MCG--6-30-15 we analysed the relationship of the fluxes
(count rates) in different energy bands following the technique of
flux--flux plots introduced by Churazov, Gilfanov \& Revnivtsev (2001)
in a study of Cygnus X--1 spectral variability, and Taylor, Uttley \&
McHardy (2003) as a model--independent tool to disentangle the main
drivers of the spectral variability in AGN.

 In Fig.~\ref{ff} we show the 3-12~keV XIS count rate from the
 45~ks light curve against the count rate in the 1--2~keV band which
 is chosen as reference (see also Vaughan \& Fabian 2004). The
 relationship between the count rates in the two bands is remarkably
 linear and leaves a clear hard offset on the y--axis, as already
 noticed by Taylor, Uttley \& McHardy (2003) and Vaughan \& Fabian
 (2004) in previous \emph{RXTE} and \emph{XMM--Newton} observations. The
 linearity of the relationship indicates that the flux variations are
 strongly dominated by changes in the normalization of a spectral
 component with constant spectral shape. On the other hand, the hard
 offset suggests the presence of a component that varies little and
 that contributes more to the hard than to the reference 1--2~keV
 band. We measure a hard offset of $a=0.48\pm 0.02$~cts/s which can be
 used to infer the contribution of the weakly variable component at
 mean flux level: since the mean count rate in the 3--12~keV band is
 $1.79\pm 0.01$~cts/s, the hard offset $a$ represents
 $26.8^{+1.3}_{-1.2}$ per cent of the 3--12~keV flux at mean flux level. This
 compares very well with the $33 \pm 6$ per cent contribution of the
 reflection component to the 3--12~keV band, as obtained from direct
 spectral fitting.

We have applied the same technique to the HXD/PIN light curve and
plotted the 14--45~keV count rate against the 1--2~keV reference
one. The relationship is much more noisy than when data from the XIS
alone were used. To reduce the scatter and decrease the PIN error
bars, we have binned the original flux--flux plot according to the
soft reference count rate so that each new point comprises three
original data points.
In the lower panel of Fig.~\ref{ff} we show the binned flux--flux plot
obtained by plotting the 14--45~keV HXD/PIN count rate against the
reference 1--2~keV one. 

A fit with a linear relationship is perfectly acceptable ($\chi^2=5.8$
for 5 dof), again indicating the presence of a clear hard offset in
the PIN band. In this case, the contribution of the weakly variable
(reflection) component to the 14--45~keV band at mean flux level is
estimated to be as high as $55\pm 17$ per cent. Again, this is in good
agreement with the contribution of the reflection component in the
14--45~keV band obtained from direct spectral fitting ($51\pm 10$ per
cent). 

The flux--flux plot analysis detailed above indicates that the spectral
variability of MCG--6-30-15 can be decomposed into two main
components: a highly variable component which varies in normalization
only but not in spectral shape, and a weakly variable one which has a
much harder spectral shape. The excellent agreement between the
contribution of the weakly variable component in the different energy
bands as inferred from the flux--flux plot analysis and that of the
reflection component as obtained by direct spectral fitting strongly
suggest that the weakly variable component is the
relativistically smeared X--ray reflection from the accretion disc.
Obviously, the highly variable component can be identified with the
other component required by spectral fitting, i.e. the power law
continuum. The flux--flux plots suggest that as the source varies, the
power law maintains the same spectral slope and only varies in
normalization.
\begin{figure}
 \begin{center}
 \includegraphics[width=0.33\textwidth,height=0.42\textwidth,angle=-90]{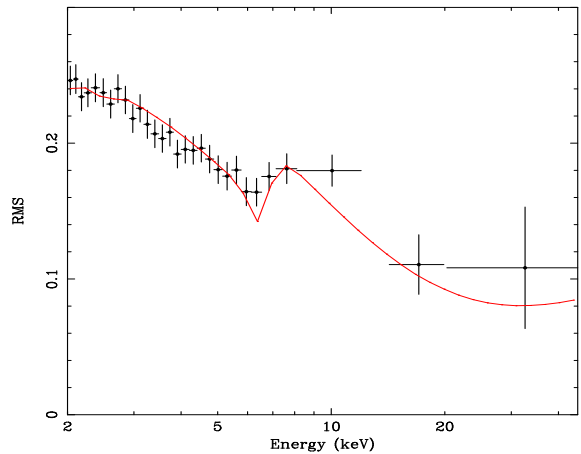}
\end{center}
\caption{
The RMS spectrum of MCG--6-30-15 on a 45~ks timescale is shown in the
2--45~keV band. The solid red line represents the theoretical RMS that
would be obtained from the best--fitting spectral model of the
time--averaged spectrum if all the variability could be explained with
a two--component model in which the power law has a constant photon
index and varies in normalization only, while the
disc reflection component is perfectly constant.}
\label{rms}
 \end{figure}

\subsection{The Flux RMS spectrum}

 A further approach to studying the spectral variability of a source
 is the RMS spectrum, which, for a chosen timescale, is the fractional
 variability seen at each energy expressed as a flux normalized root
 mean square variability function. The RMS spectrum allows us to
 quantify the fractional variability as a function of energy in a
 model--independent manner and is therefore an important piece of
 information, complementary to the flux--flux plot analysis described
 above. Such RMS spectra have previously been determined for
 MCG--6-30-15 (Matsumoto et al 2003; Fabian et al 2002; Ponti et al
 2004). However, as far as data below 10~keV are available, the
 RMS spectrum has proved to be ambiguous to model. In particular, the
 typical trend that is observed in AGN is that the fractional
 variability decreases with energy, a behaviour that can be explained
 either in terms of a power law softening at high flux levels (i.e. a
 pivoting power law) or in the framework of the two--component model
 discussed above. The new ingredient here is to include the HXD/PIN data above
 10~keV with the goal of confirming/rejecting the interpretation that
 comes out from the flux--flux plots analysis.

 Our result for a 45~ks timescale is shown in Fig.~\ref{rms}. The
 general trend is that the RMS decreases with energy with a marked
 drop at $\sim 6.4$~keV reassuringly confirming that the Fe line is
 less variable than the continuum, as already noticed in previous
 works (e.g. Vaughan \& Fabian 2004; Ponti et al 2004). The novelty of
 the \emph{Suzaku} observation is represented by the HXD/PIN data. They confirm
 the trend of lower fractional variability at higher energies but they
 lie above and not on the extrapolation of the 2--6~keV trend. 

\begin{figure}
 \begin{center}
 \includegraphics[width=0.33\textwidth,height=0.42\textwidth,angle=-90]{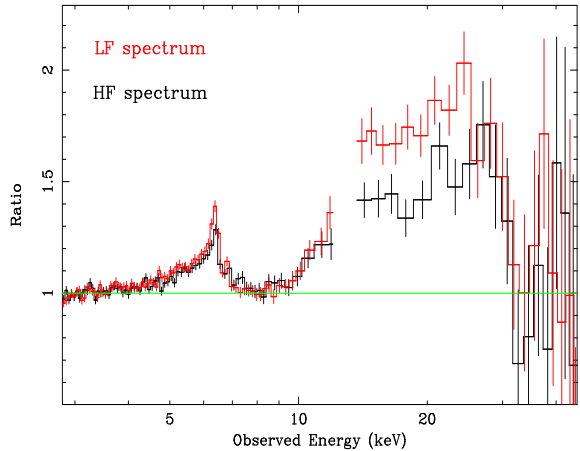}
\end{center}
\caption{Ratio plot for the HF (black) and LF (red) spectra with a
  power law model fitted ignoring the 4-7.5~keV and 14-25~keV energy
  bands. The ratio shows that the Compton hump contributes by about 80
  per cent in the LF state and by about 40--50 per cent in the HF one,
  while its contribution in the time--averaged spectrum is about 60
  per cent (see Fig.~\ref{XISandPINPLfit}). This shows that the
  reflection component is less variable than the power law and thus,
  the reflection fraction is expected to increase as the overall flux 
  decreases.}
\label{HFandLFra}
 \end{figure}

 We have theoretically modelled the RMS spectrum by considering our
 best--fitting spectral model to the time--averaged spectrum and
 reproducing the 45~ks data by allowing only the power law
 normalization to vary, i.e. enforcing a perfect two--component model.
 The result of this exercise is shown as a solid red line in
 Fig.~\ref{rms} and shows that the RMS spectrum is fully consistent
 with a two--component model in which the only variable component is a
 constant--slope power law. Some variability of the reflection
 component cannot be excluded, but its variability amplitude is
 certainly not comparable to the variability amplitude of the power law. Indeed, small
 mismatches between the model and the data could be due to some
 intrinsic variability of the reflection component which deserves more
 detailed study to be performed in future work.

\subsection{The High and Low Flux states}

Having analysed the spectral variability of MCG--6-30-15 with two
model--independent (and calibration--independent) techniques we now
consider a more direct approach through spectral fitting.  We
extracted High Flux (HF) and Low Flux (LF) from both the XIS and the
PIN spectra according to the selection criterion shown in
Fig.~\ref{totallc}. We first consider a simple power law fit in the
3--45~keV band ignoring, as done before, the Fe K and Compton hump
bands. The result for the HF and LF spectra is shown in
Fig.~\ref{HFandLFra} as a ratio plot. There is a clear evidence, both
in the XIS and in the HXD/PIN, that the reflection component
contributes more to the LF than to the HF spectrum, which again
indicates an almost constant reflection.

 We have then applied the same spectral model as for the 3--45~keV
 time--averaged (TA) spectrum to the HF and LF spectra and our results
 are reported in Table~2. Almost all parameters are
 consistent with being the same in the HF and LF spectra. In
 particular, the power law slope is consistent with being constant and
 equal to $\sim 2.25$ in the TA, HF, and LF spectra. The same is true
 for the Fe abundance, disc inclination, inner disc radius, outermost
 emissivity index $q_{\rm{out}}$ and reflector ionization state
 (although there is some indication for a lower ionization at low flux
 levels). The
 innermost emissivity index $q_{\rm in}$ appears to be steeper in the
 LF than in the HF spectrum, but the two indexes are still consistent with each
 other within the 90 per cent errors. A steeper emissivity profile in
 the LF state would produce a more extended red wing as expected in
 the framework of the light bending model where low flux states are
 characterized by a more centrally concentrated illumination of the
 inner disc (Miniutti \& Fabian 2004).

\begin{figure}
 \begin{center}
{
 \includegraphics[width=0.33\textwidth,height=0.42\textwidth,angle=-90]{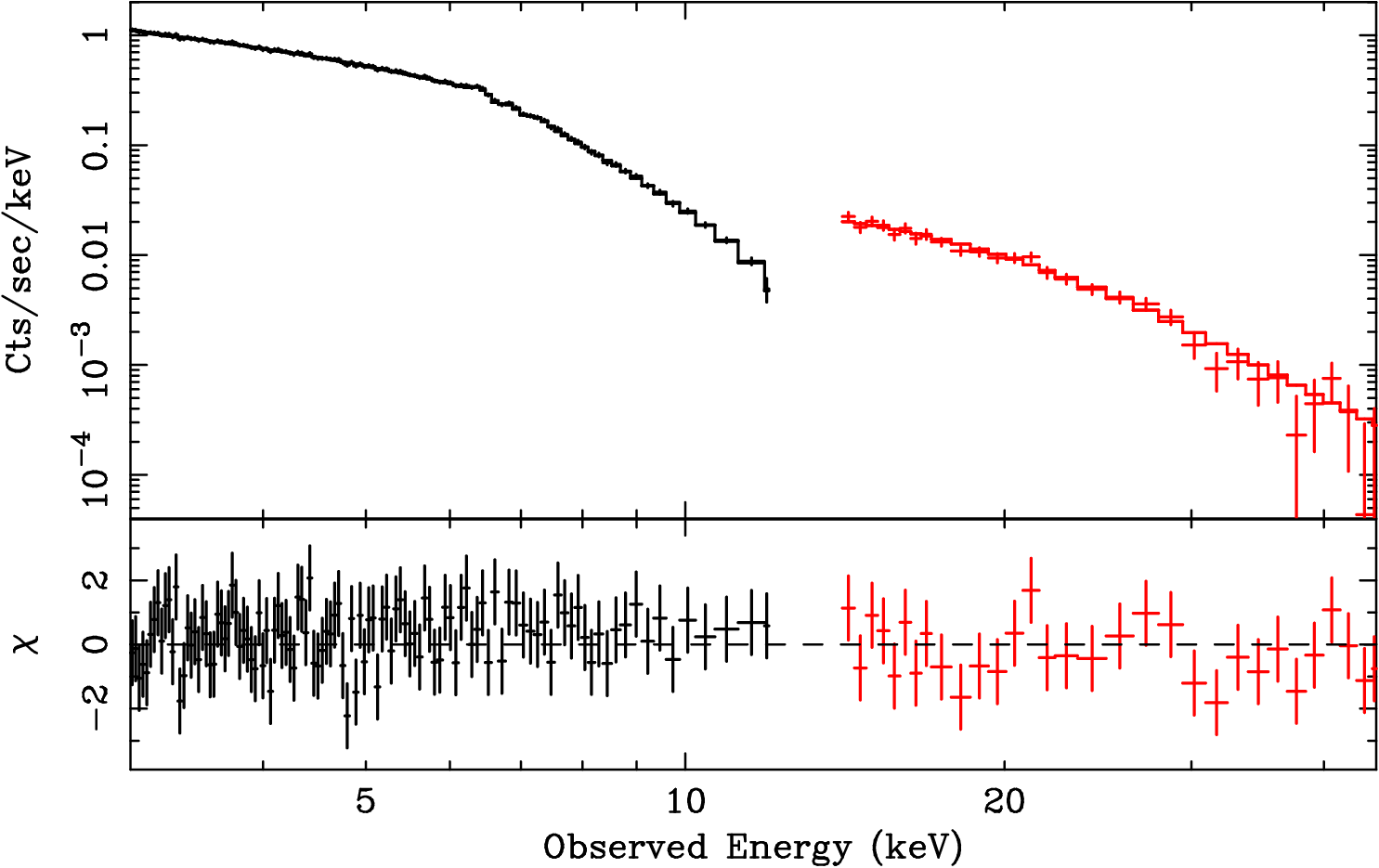}
{\vspace{0.2cm}}
 \includegraphics[width=0.33\textwidth,height=0.42\textwidth,angle=-90]{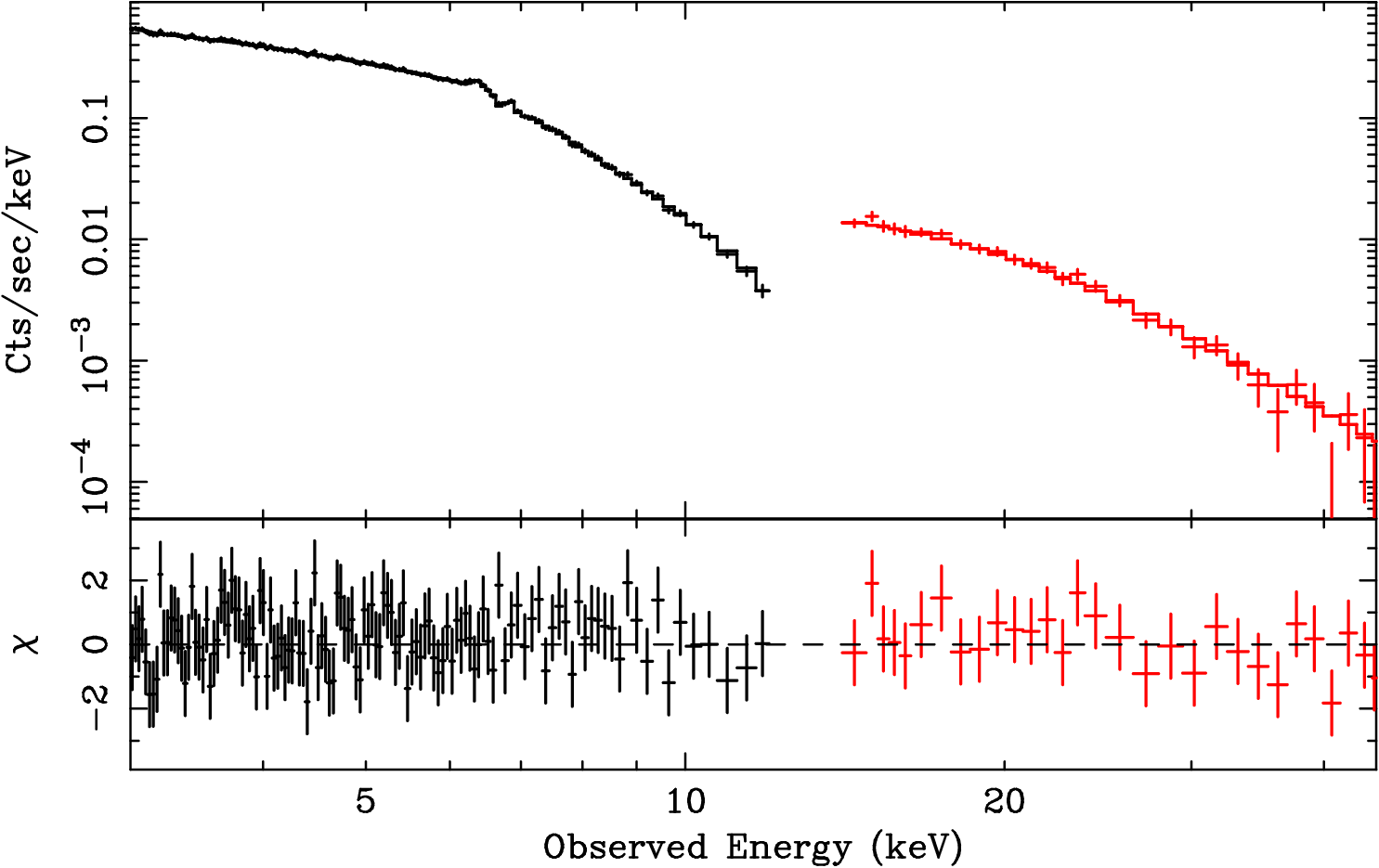}
}
\end{center}
\caption{The spectra, best--fitting model, and residuals for
  the HF (upper panel) and LF (lower panel) states.}
\label{HFandLFfit}
 \end{figure}

 On the other hand, the reflection fraction definitely appears to vary
 and it is higher in the LF spectrum ($R=4.8\pm 0.8$) than in the HF
 one ($R=2.5\pm 0.6$) as already strongly suggested by the ratio plot
 in Fig.~\ref{HFandLFra}. This result is fully consistent with the
 indication inferred from the flux--flux plot and rms spectrum
 analysis discussed above: if the reflection component is only weakly
 variable, its contribution increases in the LF states because the
 direct power law decreases while the reflection stays approximately
 constant. We point out that this is exactly the behaviour predicted
 by the light bending model (Miniutti \& Fabian 2004). The spectra,
 best--fitting models and residuals for the HF and LF states are shown
 in Fig.~\ref{HFandLFfit}. The best--fit spectral components are as in
 Fig.~\ref{Suzakumo} ( which refers to the best--fit parameters 
 to the time--averaged spectrum) with the only major difference in the
 HF/LF state being due to a higher/lower power law
 normalization.

\subsection{The difference spectrum}

\begin{figure}
 \begin{center}
 \includegraphics[width=0.33\textwidth,height=0.42\textwidth,angle=-90]{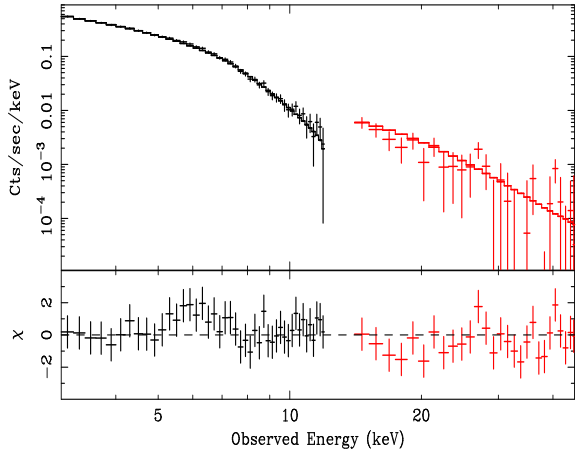}
\end{center}
\caption{The difference spectrum (HF minus LF) fitted with s power law
  absorbed by the Galactic column.}
\label{diffspec}
 \end{figure}

We also produced a difference spectrum obtained by subtracting the LF
spectrum from the HF one for the XIS and HXD/PIN. By
definition, the difference spectrum only shows the variable components
of the spectrum (modified by absorption), while any non variable
component is subtracted away. In
Fig.~\ref{diffspec} we show the residuals to a simple power law fit in
the 3--45~keV band where Galactic absorption is included. The power
law fit is very good and the power law slope is consistent with that
found in our spectral analysis ($\Gamma=2.2\pm 0.1$) but clear
positive residuals are left in a broad hump around 6~keV at the $\sim
2\sigma$ level. This may indicate either that the red wing of the
relativistic Fe line has varied in flux or that the shape of the line
is slightly different in the HF and LF spectra.

If the residuals are due to flux variability of the reflection
component, we should also see some positive residuals at the Compton
hump energy (20--30~keV) which are not seen in the HXD/PIN difference
spectrum. Some reflection variability is in fact allowed by the
results of our spectral fits of the HF and LF spectra (see Table~2)
although the reflection flux is consistent with being the same at the
two flux levels.  If the reflection flux varied but the Fe line keeps
the same shape, the main residual in the difference spectrum should be
at the energy of the blue peak of the line, around 6.4~keV, while the
residuals in Fig.~\ref{diffspec} indicate that only the red wing of
the line has varied.

The remaining possibility is that the Fe line has a different shape
(rather than a different flux) in the HF and LF spectra. This is
already hinted at by the inner emissivity index ($q_{\rm in}$, see
Table~2) which is marginally steeper in the LF state (although
consistent within the errors). If the emissivity is really different
at the two flux levels, residuals are produced in the red wing of the
line and could potentially explain those seen in the difference
spectrum (and at the same time the lack of residual at Compton hump
energies). The result is in the sense expected from the light-bending
model (see Fig.~4 of Miniutti \& Fabian 2004).

\section{Discussion}

The Suzaku observation of MCG--6-30-15 has, for the first time,
enabled the hard X-ray variability of the source to be determined. A
wide range of techniques including direct spectral fitting, flux-flux
plots and rms spectrum all show that the spectral variability on a 45~ks
timescale can be decomposed into just two broad components above
3~keV, namely a variable power-law continuum and a harder constant
component. This last component has the broad iron line and reflection
hump expected from X-ray reflection in the very innermost regions of
an accretion disc around a rapidly spinning black hole.

Just the strength and shape of the reflection component predicts a
high equivalent width iron line, much stronger than the weak narrow
emission component seen in the spectrum. The only way for such a
strong iron line to be `hidden' in the spectrum is for it to be
smeared into the broad emission structure that we observe. 

The behaviour of the power-law and reflection components is consistent
with the light bending model. In which case much of the rapid
(power-law component) variability seen in the source is due to strong
gravitational light bending as the site of the emission region changes
location within a few gravitational radii of the black hole. We assume
that the emission region is associated with some magnetic structure
produced by the strong differential motions in the accretion
disc. Note that to produce observed variability the emission region
does not need to physically move, just that the main emission site
changes position.

The broad iron line is strong in MCG--6-30-15 because both the iron
abundance is high, at around 2 times the solar value, and the large
degree of light bending around its Kerr black hole has emphasised the
reflection spectrum relative to the power-law continuum.

\section*{Acknowledgements}
We are deeply grateful to the whole Suzaku team for building,
launching, calibrating and operating the spacecraft and instruments. 
GM thanks the PPARC, ACF the Royal Society, JL the Isaac Newton
Trust, Corpus Christi College and PPARC, JC the PPARC for support.

\begin{table*}
\centering
\caption{Results of fits to the XIS2/XIS3 and PIN 
  Time--Averaged (TA), High Flux (HF) and Low Flux (LF) spectra in the
  3--45~keV band. The power law normalization ($N_{\rm PL}$) is given in units of
  $10^{-2}$~ph~cm$^{-2}$~s$^{-1}$. As for the reflection, we prefer to
  give the reflection fraction which is a more interesting measure of
  its strength. The power law and reflection fluxes are reported in
  the 3--45~keV band in units of $10^{-11}$~erg~cm$^{-2}$~s$^{-1}$ for
  completeness. The reflector ionization parameter
  ($\xi$) is in erg~cm~s$^{-1}$ and the inner disc radius ($r_{\rm{in}}$) and
  inclination ($i$) are in units of $r_g=GM/c^2$ and degrees
  respectively. The energies of the lines are in keV, their EW in
  eV, and their width is fixed to 10~eV. The narrow Fe K$\alpha$
  normalization is in units of $10^{-6}$~ph~cm$^{-2}$~s$^{-1}$. A 
  subscript $_p$ indicates that the parameter reached its min/max allowed value.}
\begin{center}
\begin{tabular}{lcccc} 
\hline \\
Parameter & TA spectrum & HF spectrum & LF spectrum \\ \\
\hline 
\\
$\Gamma$&$2.26\pm 0.04$ &  $2.25\pm 0.05$  & $2.25\pm 0.05$ \\ 
$N_{\rm PL}$ & $1.7\pm 0.1$& $2.7\pm 0.1$ & $1.2\pm 0.1$\\ \\
\hline \\
$R$& $3.8\pm 0.7$& $2.5\pm 0.6$& $4.8\pm 0.8$\\ 
$\xi$& $68\pm 31$& $70\pm 37$ & $48\pm 35$\\ 
$A_{\rm{Fe}}$& $1.9^{+1.4}_{-0.5}$& $1.8^{+1.3}_{-0.4}$&
$1.9^{+1.5}_{-0.6}$\\ 
$r_{\rm{in}}$& $1.7^{+0.4}_{-0.465p}$& $1.235^{+0.7}_{-0.0p}$&$1.7^{+0.3}_{-0.465p}$ \\ 
$i$& $38\pm 3$& $37\pm 4$& $41\pm 5$\\ 
$q_{\rm in}$& $4.4^{+0.5}_{-0.8}$& $3.6^{+0.8}_{-0.7}$ & $4.5\pm 0.6$\\ 
$q_{\rm out}$& $2.5\pm 0.3$& $2.8\pm 0.4$& $2.6\pm 0.3$\\ \\
\hline \\
$E_{\rm{FeK\alpha}}$& $6.41^{+0.03}_{-0.02}$ &  $6.40\pm 0.03$ & $6.41\pm 0.03$\\ 
$EW_{\rm{FeK\alpha}}$& $35\pm 5$& $19\pm 4$ & $44\pm 6$\\ 
$N_{\rm{FeK\alpha}}$& $8.5\pm 1.3$& $8.3\pm 1.2$ & $8.5\pm 1.3$\\ \\
\hline \\
$E_{\rm{abs_1}}$& $6.73\pm 0.04$& $6.72\pm 0.05$& $6.75\pm 0.04$\\ 
$EW_{\rm{abs_1}}$& $-(22\pm 8)$& $-(20^{+23}_{-14})$ & $-(45\pm 30)$\\ 
$E_{\rm{abs_2}}$& $7.07\pm 0.04$& $7.08\pm 0.05$ & $7.03\pm 0.04$\\ 
$EW_{\rm{abs_2}}$& $-(13\pm 8)$& $-(15\pm 10)$ & $-(28\pm 18)$\\ \\
\hline \\
$F^{\rm PL}_{3-45~\rm{keV}}$& $4.0\pm 0.2$& $6.4\pm 0.2$& $2.9\pm 0.3$\\ 
$F^{\rm REF}_{3-45~\rm{keV}}$& $4.1\pm 0.8$& $4.0\pm 0.9$& $3.8\pm 0.6$\\ \\
\hline \\
$\chi^2/dof$& 2495/2312& 1910/1848 & 1900/1925\\ \\
\hline
\end{tabular}
\end{center}
\end{table*}

\end{document}